\documentclass[twocolumn,tighten]{aastex701}
\usepackage{isotope}

\graphicspath{{figs/}}
\usepackage{amsmath}
\usepackage{graphicx}
\usepackage{multirow}
\usepackage{float}
\usepackage{xcolor}
\usepackage{enumitem}
\newif\ifdraft
\drafttrue    

\ifdraft
  
\else
  
\fi

\begin{document}

\title{Impact of Nuclear Reaction Rate Uncertainties on Type I X-ray Burst Nucleosynthesis: A Monte Carlo Study}

\author[orcid=0009-0008-8274-2521]{Qing Wang}
\affiliation{Shenzhen Key Laboratory of Research and Manufacture of High Puriny Germanium Materials and Detectors, Institute for Advanced Study in Nuclear Energy $\&$ Safety, College of Physics and Optoelectronic Engineering, Shenzhen University, Shenzhen 518060, Guangdong, People’s Republic of China}
\email{wangqing2023@email.szu.edu.cn}  

\author[orcid=0000-0002-3117-1318]{Ertao Li}
\affiliation{Shenzhen Key Laboratory of Research and Manufacture of High Puriny Germanium Materials and Detectors, Institute for Advanced Study in Nuclear Energy $\&$ Safety, College of Physics and Optoelectronic Engineering, Shenzhen University, Shenzhen 518060, Guangdong, People’s Republic of China}
\email{let@szu.edu.cn}
\correspondingauthor{let@szu.edu.cn}

\author[orcid=0000-0001-5206-4661]{Zhihong Li}
\affiliation{China Institute of Atomic Energy, Beijing 102413, People’s Republic of China}
\email{zhli@ciae.ac.cn}

\author[orcid=0000-0001-7149-814X]{Youbao Wang}
\affiliation{China Institute of Atomic Energy, Beijing 102413, People’s Republic of China}
\email{ybwang@ciae.ac.cn} 

\author[orcid=0000-0002-4911-0847]{Bing Guo}
\affiliation{China Institute of Atomic Energy, Beijing 102413, People’s Republic of China}
\email{guobing@ciae.ac.cn}

\author[orcid=0000-0001-8702-431X]{Yunju Li}
\affiliation{China Institute of Atomic Energy, Beijing 102413, People’s Republic of China}
\email{li_yunju@163.com}

\author[orcid=0000-0002-2718-9451]{Jun Su}
\affiliation{School of Physics and Astronomy, Beijing Normal University, Beijing 100875, People’s Republic of China}
\email{sujun@bnu.edu.cn}

\author[orcid=0000-0001-6724-2256]{Shipeng Hu}
\affiliation{Shenzhen Key Laboratory of Research and Manufacture of High Puriny Germanium Materials and Detectors, Institute for Advanced Study in Nuclear Energy $\&$ Safety, College of Physics and Optoelectronic Engineering, Shenzhen University, Shenzhen 518060, Guangdong, People’s Republic of China}
\email{husp@szu.edu.cn} 

\author[orcid=0009-0004-5258-9491]{Yinwen Guan}
\affiliation{Shenzhen Key Laboratory of Research and Manufacture of High Puriny Germanium Materials and Detectors, Institute for Advanced Study in Nuclear Energy $\&$ Safety, College of Physics and Optoelectronic Engineering, Shenzhen University, Shenzhen 518060, Guangdong, People’s Republic of China}
\email{2500232013@mails.szu.edu.cn}  

\author[orcid=0009-0003-3896-0269]{Dong Xiang}
\affiliation{Shenzhen Key Laboratory of Research and Manufacture of High Puriny Germanium Materials and Detectors, Institute for Advanced Study in Nuclear Energy $\&$ Safety, College of Physics and Optoelectronic Engineering, Shenzhen University, Shenzhen 518060, Guangdong, People’s Republic of China}
\email{2500231015@mails.szu.edu.cn} 

\author[orcid=0009-0008-9198-5971]{Yu Liu}
\affiliation{Shenzhen Key Laboratory of Research and Manufacture of High Puriny Germanium Materials and Detectors, Institute for Advanced Study in Nuclear Energy $\&$ Safety, College of Physics and Optoelectronic Engineering, Shenzhen University, Shenzhen 518060, Guangdong, People’s Republic of China}
\email{2500232005@mails.szu.edu.cn} 

\author[orcid=0009-0009-2232-4317]{Lei Yang}
\affiliation{Shenzhen Key Laboratory of Research and Manufacture of High Puriny Germanium Materials and Detectors, Institute for Advanced Study in Nuclear Energy $\&$ Safety, College of Physics and Optoelectronic Engineering, Shenzhen University, Shenzhen 518060, Guangdong, People’s Republic of China}
\email{2310456020@email.szu.edu.cn}

\author[orcid=0000-0003-3233-8260]{Weiping Liu}
\affiliation{Department of Physics, Southern University of Science and Technology, Shenzhen 518055, Guangdong, People’s Republic of China}
\email{wpliu@ciae.ac.cn}

\begin{abstract}
Type I X-ray bursts are thermonuclear flashes on the surface of accreting neutron stars, involving hundreds of nuclei and thousands of reactions with larger uncertainties in reaction rate. To investigate the impact of nuclear reaction rate uncertainties on type I X-ray burst nucleosynthesis, comprehensive Monte Carlo simulations were performed with temperature-independent and -dependent variations in reaction rates using the REACLIB and STARLIB libraries, respectively.
A total of 1,711 $(p, \gamma)$, $(p, \alpha)$, $(\alpha, p)$, and $(\alpha, \gamma)$ reaction rates are varied simultaneously along with their inverse reactions via detailed balance. 
For the first time, it has been found that Monte Carlo sampling with larger perturbations to these reaction rates may lead to multi-peak abundance distributions for certain isotopes, such as $^{64}$Zn and $^{55}$Co. These multi-peak structures arise not only from coupled reactions but also from single reactions in some cases.
Our studies also confirm previously identified key reactions and provide more robust lists that deserve priority consideration in future studies.
\end{abstract}


\section{\label{sec:Introduction}Introduction} 
Type I X-ray bursts (XRBs) are the most frequent thermonuclear explosions observed in our Galaxy  ~\citep{Cyburt2016,Parikh2013}. They are powered by unstable hydrogen and helium burning on the accreted envelope of neutron stars in low mass X-ray binary systems (LMXBs) \citep{Hansen1975,Woosley1976,Joss1977,Cyburt2010}. In the regime of combined ignition of H/He, XRBs are dominated by the $3 \alpha$ reaction, the $\alpha p$-process (a suite of $(\alpha, p)$ and $(p,\gamma)$ reactions) and the $rp$-process (a sequence of rapid proton captures and $\beta^+$ decays)~\citep{Fisker2008,Wallace1981,Woosley2004,Schatz2001,Jos2010,Parikh2013}. Under the most favorable conditions (high hydrogen content in the accreted matter, low metallicity, and high accretion rate, etc.), reactions can proceed up to and terminate in the SnSbTe cycle ~\citep{Schatz2001,Cyburt2010}. These nuclear processes involve hundreds of nuclear species, from stable isotopes to proton drip-line nuclei. Thousands of nuclear reaction rates have not been precisely determined with large uncertainties~\citep{Cyburt2016}. To explore these uncertainties in XRB nucleosynthesis models, extensive investigations have been conducted. 

Varying reaction rates individually is a classical method for quantifying the numerical sensitivity of observables to specific reactions ~\citep{Schatz2016} in XRB models. In particular, ~\citet{Cyburt2016} investigated the impact of uncertainties in $(p, \gamma)$, $(\alpha, \gamma)$, and $(\alpha, p)$ reaction rates by individually varying each, first screening sensitive reactions in the onezone model and then recalculating the important ones in the fully self-consistent 1D multizone model. They found some reactions that significantly affect the predictions of light curves and burst ashes.~\citet{Lam2022a} carried out a sensitivity study of individual reactions to examine their influence on photospheric radius expansion of XRBs, and found that the observables are more sensitive to some $\alpha$-capture reactions.

Nevertheless, it has been claimed~\citep{Roberts2006} that due to the fact that many different channels are coupled in nucleosynthesis, traditional sensitivity studies, in which only one reaction is varied, and all the others remain constant, cannot adequately address all the important correlations and may lead to wrong (or at least biased) conclusions. Monte Carlo studies have the advantage that all correlations between reactions are taken into account and that uncertainties can be used to properly and quantitatively propagate to predicted observables~\citep{Schatz2016}. They have been performed for several other astrophysical scenarios: Big Bang Nucleosynthesis (BBN) ~\citep{Iliadis2020}, $\nu p$-process~\citep{Nishimura2019}, $s$-process~\citep{Cescutti2018}, $r$-process~\citep{Mumpower2016}, weak $r$-process~\citep{Bliss2020, Psaltis2022}, $i$-process ~\citep{Denissenkov2021}, $\gamma$-process~\citep{Rauscher2016}, weak $rp$-process ~\citep{Psaltis2025}, AGB stars~\citep{Fields2016} and massive stars~\citep{Fields2018}. For XRBs,~\citet{Parikh2008} systematically varied 3,500 nuclear reactions individually and simultaneously based on temperature and density profiles using onezone post-processing models.

In some of these and similar studies like~\citet{Parikh2008}, varying reaction rates by a constant multiplicative factor assumes that the true, unknown reaction rate differs from the recommended rate by that factor over all temperatures (a temperature-independent Monte Carlo method). In other words, this method treats the rate uncertainties as temperature-independent. Although this method may lead to either underestimation or overestimation of the reaction rate at different stellar temperatures~\citep{Fields2016, Fields2018}, it has been adopted in subsequent studies to explore the uncertainties of theoretical reaction rates in weak $r$-process nucleosynthesis, e.g.~\citet{Bliss2020, Psaltis2022}. In the implementation of this Monte Carlo method,~\citet{Parikh2008} and \citet{Bliss2020} both assumed that reaction rate uncertainties are roughly a factor of 10 according to theoretical model estimates without a rigorous statistical meaning. 
However, they adopted different statistical interpretations of this same physical range.~\citet{Parikh2008} effectively mapped it to a lognormal distribution corresponding 
to a 95\% confidence interval (C.I.). In contrast,~\citet{Bliss2020} adopted a lognormal distribution whose 68\% C.I. spans a factor of 10, representing a more conservative 
treatment.

In the statistical analysis of the final results, both~\citet{Parikh2008} and  ~\citet{Bliss2020} reported the 95\% C.I. abundance uncertainties. However, the treatment of uncertainties in~\citet{Bliss2020} is more conservative than that of~\citet{Parikh2008},  due to the broader lognormal distribution adopted for the reaction rates.

In reality, the uncertainties of an individual reaction rate are often temperature-dependent when based on experimental constraints, usually with larger uncertainties at lower temperatures. For XRB nucleosynthesis, although most reaction rates are calculated using theoretical models, some key reactions are experimentally determined, making the use of realistic uncertainties essential in Monte Carlo simulations. A more robust approach incorporates temperature-dependent uncertainties, as provided by the STARLIB library~\citep{Sallaska2013}. STARLIB provides reaction rate probability density functions in the form of lognormal distributions, with factor uncertainties corresponding to the 68\% C.I. (or a coverage probability of 68\%)~\citep{Sallaska2013,Iliadis_2015}, which enables Monte Carlo simulations for more realistic nucleosynthesis calculations (hereafter referred to as a temperature-dependent Monte Carlo method; see also~\citealt{Longland2012,Fields2016,Fields2018, Psaltis2025}). In the statistical analysis of the final results, there is also no uniform standard adopted in STARLIB-based studies, with some works reporting the 95\% C.I.~\citep[e.g.,][]{Fields2018} and others reporting the 68\% C.I.~\citep[e.g.,][]{Psaltis2025}.
 
Motivated by the above considerations and to address gaps in previous studies, we conduct a comprehensive Monte Carlo study to assess how different treatments of reaction rate uncertainties affect XRB nucleosynthesis. In particular, we focus on the variations in final ash compositions, as well as key reactions shaping the burst ashes. We first examine the impacts of the two representative sampling schemes in the temperature-independent Monte Carlo method with the REACLIB library~\citep{Cyburt2010}, analyzing their influence on final abundances and the identification of key reactions. We then extend this analysis by implementing a temperature-dependent Monte Carlo method based on the STARLIB library. Together, the three Monte Carlo schemes are compared to evaluate the sensitivity of XRB nucleosynthesis to the treatments of reaction rate uncertainties.

This paper is organized as follows: Section~\ref{sec:Method and Input Physics} provides a detailed description of the models and methods used in our studies. Section~\ref{sec:Result} presents the results of our simulations. Finally, Section~\ref{sec:Discussion and Conclusion} discusses the implications of our findings and draws conclusions.

\section{\label{sec:Method and Input Physics}Models and Methods}
\subsection{\label{subsec:Models}Models: Temperature and Density profiles}
\begin{figure*}[!htbp]
	\centering
	\includegraphics[width=\textwidth]{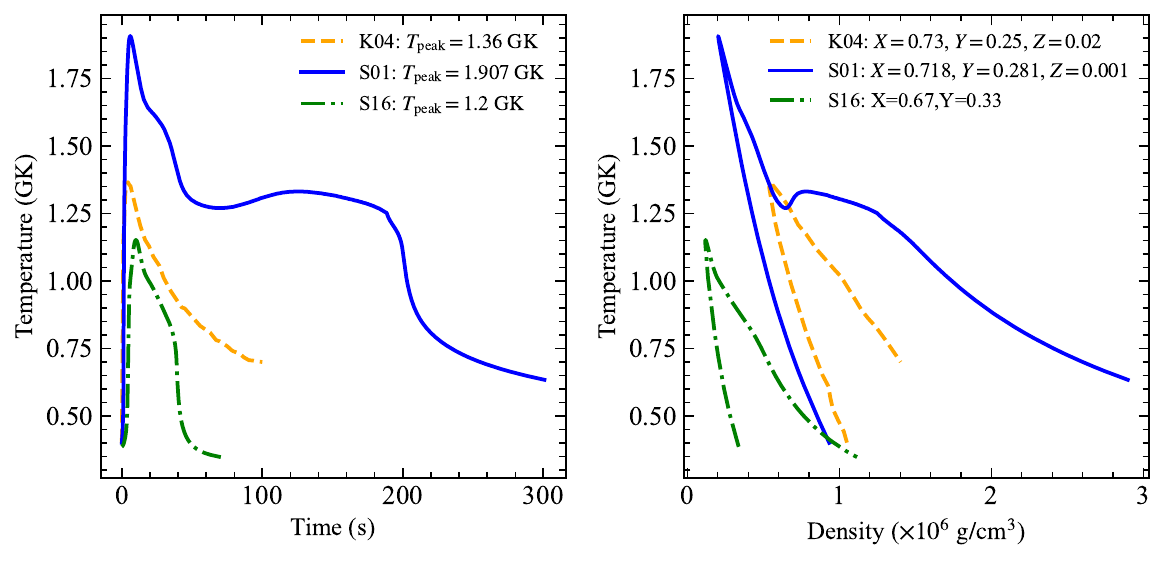}
	\caption{\label{fig:profiles}    
    (Left) Time evolution of temperature (in units of $10^9~\mathrm{K}$) for the XRB Models, K04 (orange dashed line)~\citep{Koike2004}, S01 (blue solid line)~\citep{Schatz2001}, and S16 (green dot-dashed line)~\citep{Cyburt2016}. (Right) Same as left panel, but for temperature vs. density (in units of $10^6~\mathrm{g/cm^{3}}$). The peak temperature and initial metallicity for each model are also listed.
 }
\end{figure*}

We used three different XRB models, K04~\citep{Koike2004},  S01~\citep{Schatz2001}, and  S16~\citep{Cyburt2016}. Their temperature and density profiles are shown in Figure~\ref{fig:profiles}.

Model K04 has a maximum temperature of 1.36 GK, with densities ranging between ($0.54$ - $1.44) \times 10^6~\mathrm{g/cm^{3}}$, and a burst duration of approximately 100 s. Its initial composition follows~\citet{Parikh2008} (based on~\citet{Koike2004}), with $X=0.73$, $Y=0.25$, and $Z=0.02$, where $Z$ is scaled according to the solar abundances~\citep{Grevesse1998}.

Model S01 has the highest maximum temperature of $1.907$ GK among the considered models, and lasts for about $300\,\mathrm{s}$. Unlike previous studies ~\citep{Parikh2008} that scaled densities from K04, we adopt the actual density profile for this model. The initial composition is $X=0.718$, $Y=0.281$, and $Z=0.001$. Following ~\citet{Parikh2008}, we assume that all metals initially are in the form of $^{14}$N, as described by~\citet{Woosley2004}.

Model S16 represents a burst characterized by a lower maximum temperature of $1.2$~GK. The trajectory was calculated using the onezone model. Its ignition conditions are chosen from the multizone model so that the resulting luminosity and final composition closely resemble those of the multizone calculation. The selected ignition conditions correspond to a temperature of 0.386 GK, a pressure of $1.73 \times 10^{22}$~$\mathrm{erg/cm^{3}}$, and hydrogen and helium mass fractions of 0.51 and 0.39, respectively, as reported by ~\citet{Cyburt2016}. Instead of using $X=0.51$ and $Y=0.39$ as the initial composition in this work, we adopted $X=0.67$ and $Y=0.33$, following ~\citet{Zhang2024} with zero metallicity, taken from~\citet{Schatz2001}.

\subsection{\label{subsec:Method} Nucleosynthesis Methods}
Nucleosynthesis studies of XRBs are computationally challenging due to short characteristic timescales and high peak temperatures. The large number of nuclear species and extensive reaction networks (several thousand nuclear reactions) required to describe the $rp$-process and $\alpha p$-process in H- and He-rich mixtures accreted onto the surface of a neutron star further increase the computational complexity. Since full reaction networks coupled to multizone hydrodynamic models remain computationally prohibitive~\citep{Parikh2013}, onezone post-processing~\citep{Parikh2008} or onezone~\citep{Cyburt2016} calculations are often adopted as practical alternatives for comprehensive studies. 

In this work, the onezone post-processing method is adopted using the WinNet code. The NucNet Tools~\citep{Meyer2007} and the SkyNet code~\citep{Lippuner2017} are also used to cross-check the results. The network consists of 686 nuclides, from hydrogen to $^{136}$Xe, and includes more than 8,000 reactions. Among these, 1,711 forward reactions—$(p, \gamma)$, $(p, \alpha)$, $(\alpha, p)$ and $(\alpha, \gamma)$—are varied, and the corresponding inverse rates are calculated via detailed balance, while all other reactions remain unchanged and are adopted from REACLIB v2.2~\citep{Cyburt2010}. In particular, the $3\alpha$ rate is not varied, as varying it by a factor of 10 would lead to unphysical results~\citep{Parikh2008}. For the temperature-independent Monte Carlo method, these reactions are randomly sampled using lognormal multiplicative factors around their REACLIB median (recommended) rates. For the temperature-dependent Monte Carlo method, they follow the median rates and associated uncertainties from STARLIB v610~\citep{Sallaska2013}. Thus, each of the reactions is independently perturbed according to a lognormal distribution based on its REACLIB or STARLIB rate. A detailed description of the sampling procedure is presented in Section~\ref{subsec:Monte Carlo Sampling Strategy}.

Although~\citet{Parikh2008} conducted preliminary Monte Carlo studies with 1,000 trials and found that increasing the number of trials to 10,000 did not reveal additional correlations, we further extended our simulations to 100,000 trials per model and run. This allows investigation of rare tail events and improves statistical reliability.

After the nucleosynthesis evolutions are completed, all radioactive species with half-lives shorter than $1~\mathrm{h}$ are allowed to decay into their respective stable daughter nuclides, following ~\citet{Parikh2008,Psaltis2025}. Instead of using absolute isotopic abundances, we consider relative abundances, i.e., abundances from the Monte Carlo trials normalized to baseline abundances. For the temperature-independent Monte Carlo method, REACLIB v2.2 is used to calculate the baseline abundances $X_{\rm base}$. For the temperature-dependent Monte Carlo method, REACLIB v2.2, with the 1,711 forward rates replaced by the median rates of STARLIB v610, is used to compute $X_{\rm base}$. Note that all inverse rates are also calculated via detailed balance. 

\subsection{\label{subsec:Monte Carlo Sampling Strategy}Monte Carlo Sampling Methods}

\subsubsection{\label{subsubsec:Temperature-dependent Monte Carlo}Temperature-Dependent Monte Carlo Method}

Classical reaction rates are often reported with sharp upper and lower limits, lacking a rigorous statistical meaning~\citep{Longland2010}. For example, the NACRE evaluation~\citep{Angulo1999} provides recommended, lower, and upper rates as fixed boundaries, without specifying any coverage probability.  

The STARLIB library provides the median nuclear reaction rate, $r_{med}$, and the associated factor uncertainty, $f.u.$ (68\% C.I.), at temperature ranges from 0.001 GK to 10 GK~\citep{Sallaska2013}. It should be noted, however, that for reactions without available experimental data, STARLIB assigns a fixed uncertainty factor of 10 across all temperatures, based on the Hauser–Feshbach statistical model. All rates in STARLIB follow a lognormal probability distribution at each temperature $T$.
\begin{equation}
	f(r)=\frac{1}{\sqrt{2\pi}\sigma r}\, 
	e^ {-(ln(r) - \mu)^2/(2 \sigma^2)},~\mathrm{for~0<r<\infty},
\end{equation} 
where $r$ is the reaction rate. The lognormal function has two parameters $\mu$ and $\sigma$. Following~ \citet{Longland2010,Sallaska2013}, $r_{med}$ is related to the location parameter $\mu$, $r_{med}= e^\mu$. $f.u.$ is related to the spread parameter $\sigma$, $f.u.= e^\sigma$.

In each Monte Carlo trial $i$, a random variation factor $p_{ij}$ is assigned to each reaction rate~\citep{Longland2012,Sallaska2013,Fields2016}:
\begin{equation}\label{eq:tdmc}
	r_{ij} = e^{\mu} (e^{\sigma})^{p_{ij}} = r_{\rm med} \, (f.u.)^{p_{ij}},
\end{equation}
where $p_{ij}$ is drawn from a normal distribution with a mean of zero and a standard deviation of unity for the reaction rate $j$. Although $p_{ij}$ could in principle be temperature-dependent, we adopt the simplest way in which $p_{ij}$ is constant across temperatures, which simplifies the sampling procedure and provides a good approximation, as demonstrated by~\citet{Longland2012}. When the rate variation factor $p_{ij}=0$, the median rate $r_{med}$ can be recovered. Since the factor uncertainty $f.u.$ varies with temperature, a sampled rate distribution maintains the temperature dependence of the rate uncertainty. For more detailed discussions of the sampling procedure used in Monte Carlo nucleosynthesis studies, the reader is referred to~\citet{Longland2012}.

\subsubsection{\label{subsubsec:Temperature-Independent Monte Carlo}Temperature-Independent Monte Carlo Method}
In the temperature-independent Monte Carlo studies, all reaction rates are also sampled from lognormal distributions, but with a constant factor uncertainty applied uniformly across all temperatures, as in Equation~\ref{eq:tdmc}. 
Specifically, we adopt two representative factor uncertainties: $f.u.=\sqrt{10}$ following~\citet{Parikh2008}, and $f.u.=10$ following~\citet{Bliss2020}. 
These factor uncertainties correspond to 68\% C.I.s. 
The sampling exponent $p_{ij}$ is drawn from a standard normal distribution, as in the temperature-dependent Monte Carlo method.

\begin{figure*}[!htbp]
	\centering
	\gridline{
		\fig{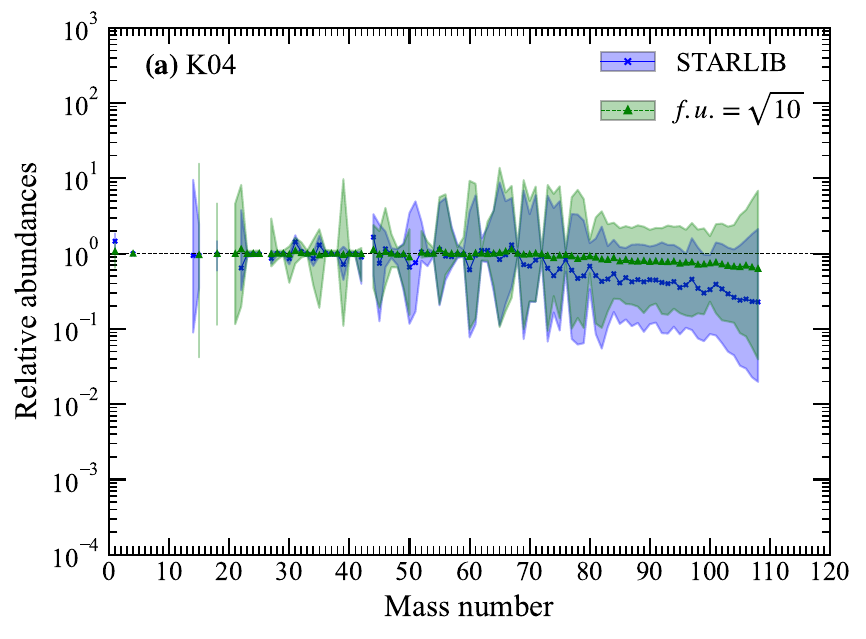}{0.5\textwidth}{}
		\fig{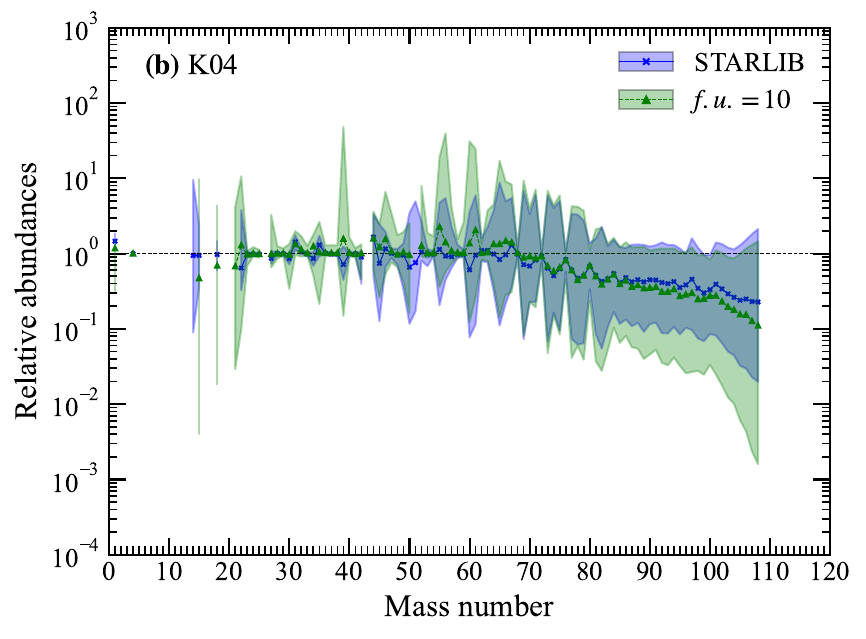}{0.5\textwidth}{}            
	}
	\vspace{-30pt} 
	\gridline{
		\fig{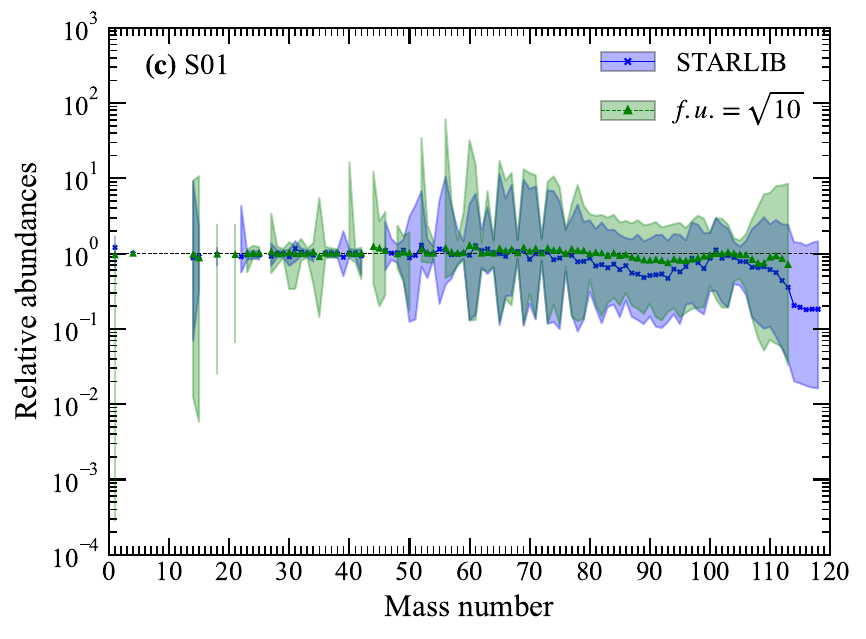}{0.5\textwidth}{}
		\fig{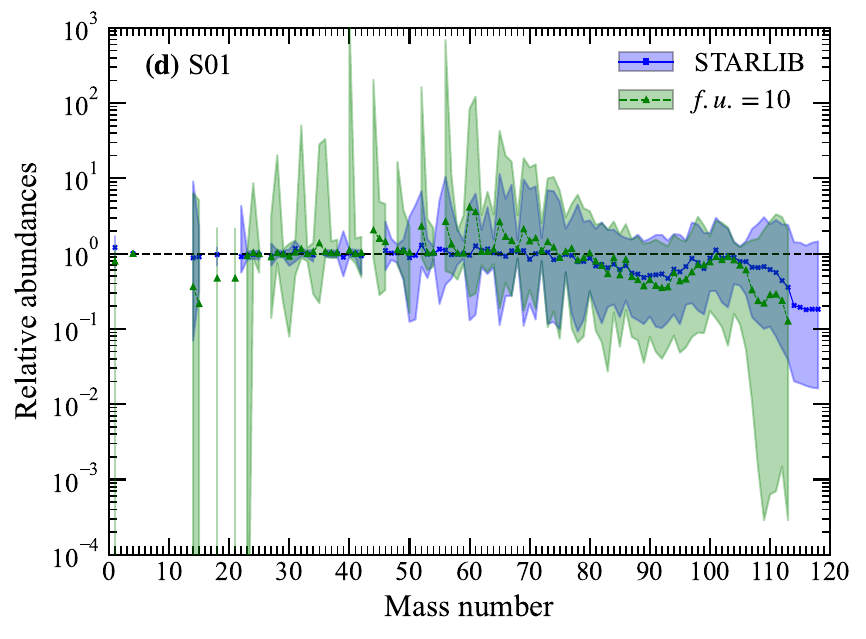}{0.5\textwidth}{}    
	}
	\vspace{-30pt} 
	\gridline{
		\fig{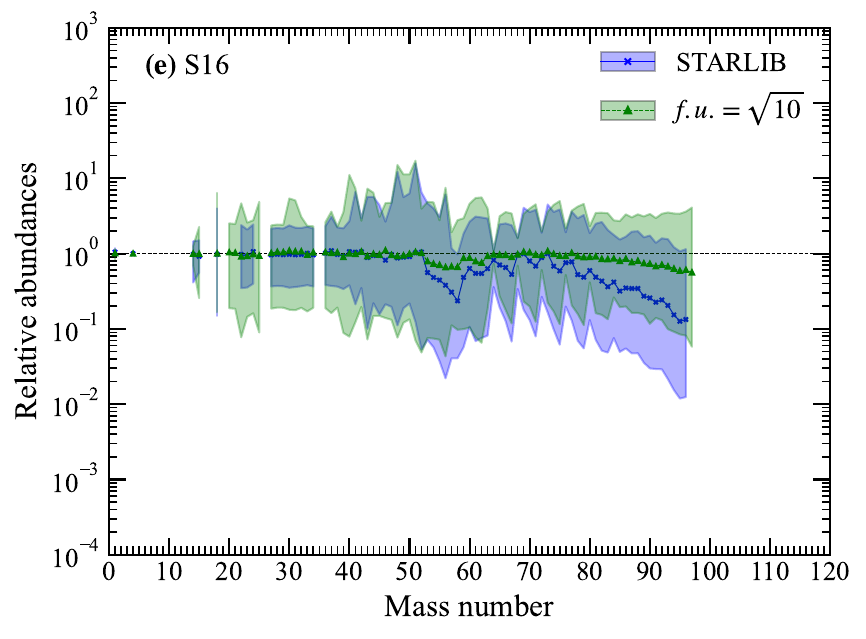}{0.5\textwidth}{}
		\fig{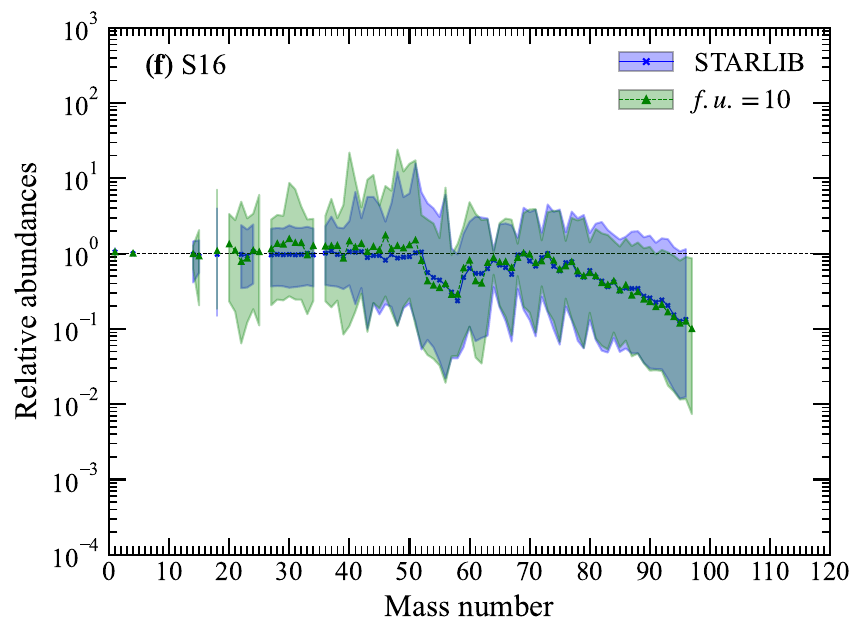}{0.5\textwidth}{}
	}
	\vspace{-20pt}
	\caption{\label{fig:TDMC}Comparison of final abundances ($X_{\rm base} \geq 10^{-6}$) between temperature-dependent Monte Carlo simulations (STARLIB) and temperature-independent simulations with $f.u.=\sqrt{10}$ (left panels) and $f.u.=10$ (right panels) for Models K04, S01, and S16 (top to bottom). Median abundances are shown as lines and symbols, with uncertainties shown as shaded bands (see legend for details): 95\% C.I. for $f.u.=\sqrt{10}$, 68\% C.I. for $f.u.=10$. For STARLIB, abundance uncertainties correspond to the 68\% C.I., same as $f.u.$}
\end{figure*}

\subsection{\label{sec:Identiﬁcation of key reactions}Identification of Key Reactions}
In both temperature-dependent and temperature-independent Monte Carlo simulations, the impact of reaction rate uncertainties on final abundances of isotopes can be calculated using correlation coefficients. Previous studies employed the Pearson correlation coefficient to capture linear relationships~\citep{Nishimura2019,Rauscher2016}, while the Spearman rank-order correlation (SROC) captures monotonic trends, including nonlinear ones~\citep{Fields2016,Fields2018,Bliss2020,Psaltis2022}. More recently, mutual information (MI) has been introduced to quantify correlations that are neither linear nor monotonic~\citep{Iliadis2020,Psaltis2025}. In this work, we focus on SROC as it provides a clear measure of both the strength and direction of these monotonic dependencies.

For the $J = 1711$ reactions, each reaction has $N = 10^5$ pre-generated Monte Carlo samples, forming the matrix $p = \{p_{ij}\} \in \mathbb{R}^{N\times J}$, with the corresponding final abundances of isotopes $X = \{x_{ik}\} \in \mathbb{R}^{N\times K}$. 

Each column of $p$ and $X$ is replaced by its rank values, resulting in the matrices $p_\mathrm{rank} \in \mathbb{R}^{N\times J}$ and $X_\mathrm{rank} \in \mathbb{R}^{N\times K}$. Each ranked column is then standardized using its column-wise mean and sample standard deviation:
\begin{equation}
	\tilde{p} = \frac{p_\mathrm{rank} - \bar{p}}{s_p}, \qquad
	\tilde{X} = \frac{X_\mathrm{rank} - \bar{X}}{s_X},
\end{equation}
where $\bar{p}$ and $\bar{X}$ denote the column-wise means, $s_p$ and $s_X$ are the corresponding standard deviations.
The Spearman rank correlation matrix $R_s \in \mathbb{R}^{J\times K}$ is obtained as
\begin{equation}
	R_s = \frac{\tilde{p}^{\top}\tilde{X}}{N-1}.
\end{equation}

The element $(j,k)$ of the matrix $R_s$ represents the Spearman correlation coefficient, denoted as $r_s$, between the reaction $j$ and the isotope $k$. A value of $r_s = +1$ indicates a perfectly monotonically increasing relationship, while $r_s = 0$ indicates the absence of a monotonic relationship. In contrast, $r_s = -1$ indicates a perfectly monotonically decreasing relationship. We have confirmed that these results align with those obtained using the \texttt{scipy.stats.spearmanr} function~\citep{2020SciPy-NMeth}, following time-consuming loop calculations for all pairs of isotopes and reactions.

Spearman correlation coefficients alone cannot reliably identify the most important reactions for a given isotope~\citep{Parikh2008,Bliss2020}, as strong correlations may exist even when the reaction rates are known to high precision and the resulting abundance variations are small. To identify key reactions, we considered both the magnitude of the isotopic abundance variations and the associated Spearman correlation coefficients. Specifically, a reaction is regarded as important if the final abundance variation is at least a factor of two and the absolute Spearman correlation coefficient satisfies $|r_s| \ge 0.5$, following~\citet{Parikh2008}.

\section{\label{sec:Result}Results}

In this work, we focus on isotopes with baseline abundances $X_{\rm base} \geq 10^{-6}$. For temperature-independent simulations, abundance uncertainties are estimated at the 95\% C.I. for $f.u.=\sqrt{10}$ and at the 68\% C.I. for $f.u.=10$. For temperature-dependent simulations using STARLIB, abundance uncertainties are estimated at the 68\% C.I., consistent with the confidence interval for $f.u.$, following~\citet{Psaltis2025}.

Table~\ref{tab:Key Reaction Table} in the Appendix shows the key reactions that affect isotopic abundances for Model K04, S01, and S16 in different sampling schemes. There are three Hot-CNO breakout reactions: $^{14}\mathrm{O}(\alpha, p)^{17}\mathrm{N}$, $^{15}\mathrm{O}(\alpha, \gamma)^{19}\mathrm{Ne}$, $^{18}\mathrm{O}(\alpha, p)^{21}\mathrm{Na}$; the remaining are $(p, \gamma)$ reactions. More discussions will be provided in section~\ref{sec:Key Reactions}. Key reactions are identified using the following criteria: 

\begin{enumerate}[label=(\roman*)]
    \item $|r_s| \ge 0.50$.
    \item Abundance variations are at least a factor of 2, defined as the ratio between the maximum and minimum isotopic abundances within the estimated uncertainty ranges mentioned above for different sampling schemes (see  Table~\ref{tab:Uncertainty Table} in the Appendix).
\end{enumerate}

\subsection{\label{sec:Relative abundances}Relative Abundances} 
Figure~\ref{fig:TDMC} compares the final abundances obtained with STARLIB to those obtained using $f.u.=\sqrt{10}$ and $f.u.=10$. Median abundances and their uncertainties are shown as functions of mass number.

As seen in the first column of Figure~\ref{fig:TDMC}, the baseline abundances with $X_{\rm med} \ge 10^{-6}$ extend up to $A = 118$ for Model S01 when using STARLIB, whereas they reach only $A = 112$ for $f.u.=\sqrt{10}$ using REACLIB. For Model K04, both approaches yield the same upper mass number. For Model S16, the corresponding limits are $A = 96$ and $A = 97$, respectively. These differences reflect how the median (recommended) rates of STARLIB and REACLIB differ and affect XRB nucleosynthesis from high to low peak temperatures.

The second column shows that median abundances exhibit similar trends at high mass numbers for STARLIB and $f.u. = 10$ across the three models, with a clear decreasing trend for Models K04 and S16.

The abundance uncertainties for $f.u. = 10$ are generally larger than those for STARLIB and $f.u. = \sqrt{10}$. In particular, for Model S01, as shown in Figure~\ref{fig:TDMC}(d), the abundance variations within estimated uncertainty ranges for the mass number $A = 1, 14, 15, 18, 21, 23$ (corresponding to isotopes $^{1}$H, $^{14}$N, $^{15}$N, $^{18}$F, $^{21}$Ne and $^{23}$Na) are noticeably overestimated, exceeding a factor of 10,000. Some mass numbers with small uncertainties reveal the existence of the waiting-point nuclides $^{68}$Se, $^{72}$Kr, $^{76}$Sr~\citep{vanWormer1994,Woosley2004,Schatz2006}, and $^{80}$Zr~\citep{Parikh2013}. Moreover, $^{64}$Ge~\citep{Woosley2004,Schatz2006,Zhou2023} is identified as an additional waiting-point nuclide in Models S01 and S16.

\subsection{\label{sec:Key Reactions}Key Reactions}
\begin{figure*}[!htp]
	\centering
	\includegraphics[width=\textwidth]{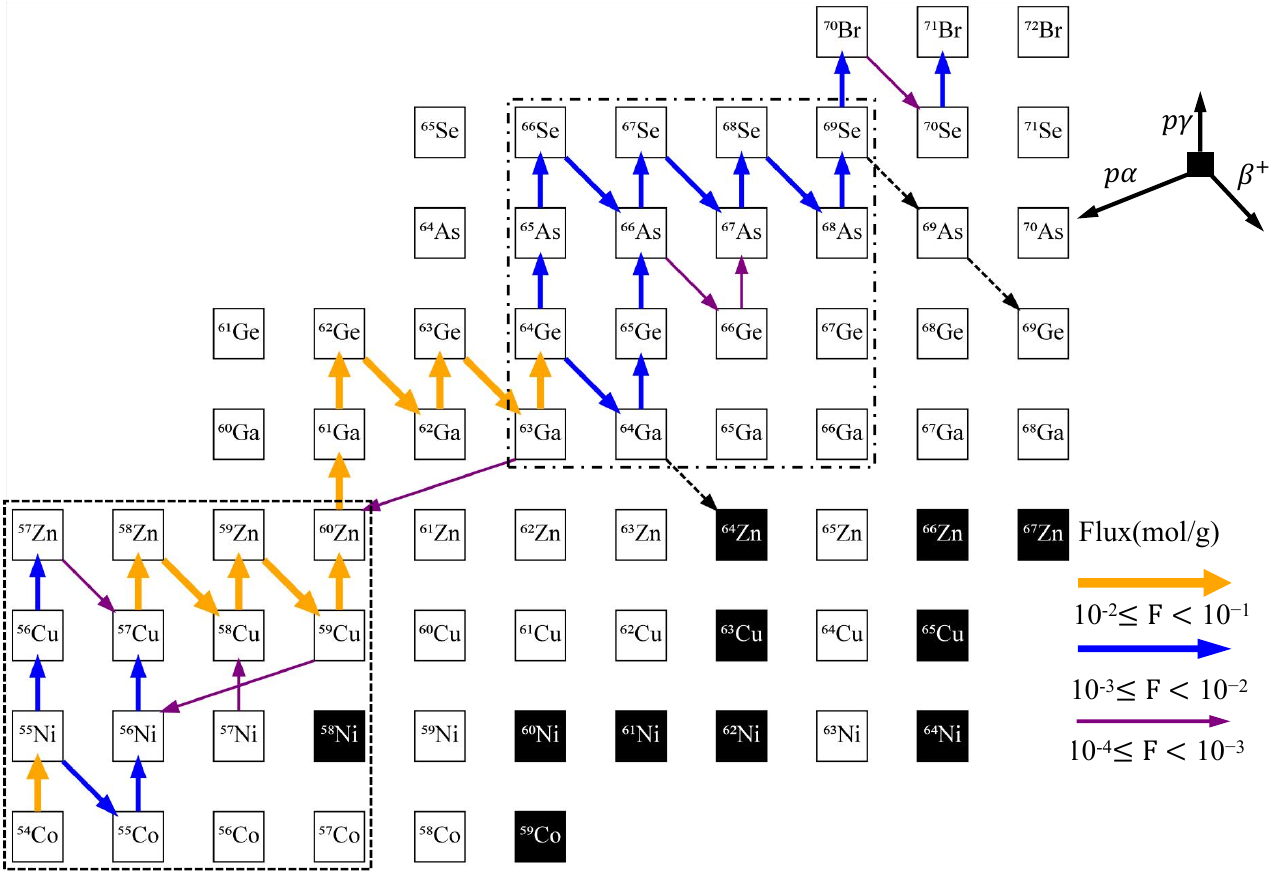}
	\caption{\label{fig:flow_standrates}Integrated net fluxes at the end of the trajectory for Model K04 using standard rates (REACLIB v2.2). The $\beta^+$-decays of $^{64}$Ga ($T_{1/2}= 2.627\,\mathrm{min}$), $^{69}$Se ($T_{1/2}= 27.4\,\mathrm{s}$), and $^{69}$As ($T_{1/2}= 15.2\,\mathrm{min}$) after the end of the trajectory are shown as black dashed arrows. For clarity, fluxes below $10^{-4}\,\mathrm{mol/g}$ and certain channels (e.g., $(\alpha,\gamma)$) are not displayed. The stable nuclides are indicated as black-filled squares. Flux strengths are indicated by arrow sizes and colors, with the corresponding legend shown on the right.}
\end{figure*}

In this section, we present how different sampling schemes affect the identification of key reactions and the isotopic abundance distributions in Model K04. In particular, we focus on the abundance distributions of $^{69}$Ge, $^{55}$Co, and $^{64}$Zn and the reactions that affect them. Figure~\ref{fig:flow_standrates} illustrates their locations in the reaction network. The abundance distributions of $^{55}$Co and $^{64}$Zn become bimodal under certain sampling schemes, highlighting a main finding of this study.

\begin{figure}[!htbp]
\centering
\includegraphics[width=0.40\textwidth]{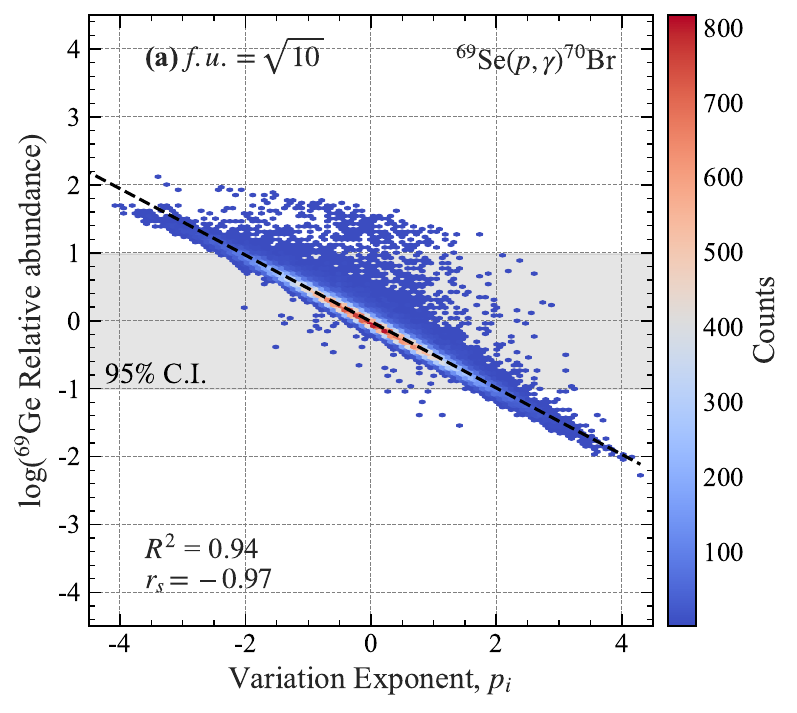}\\
\includegraphics[width=0.40\textwidth]{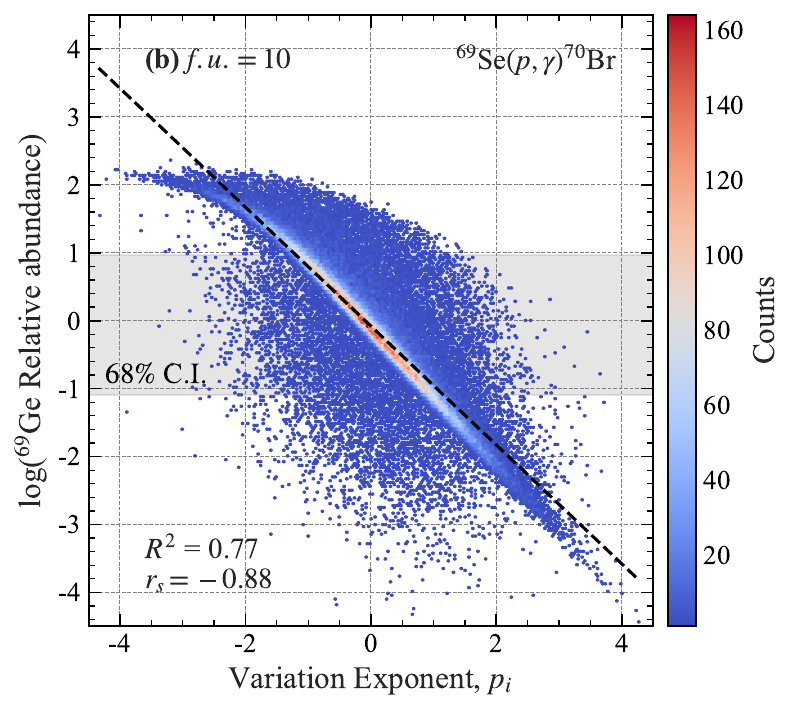}\\
\includegraphics[width=0.40\textwidth]{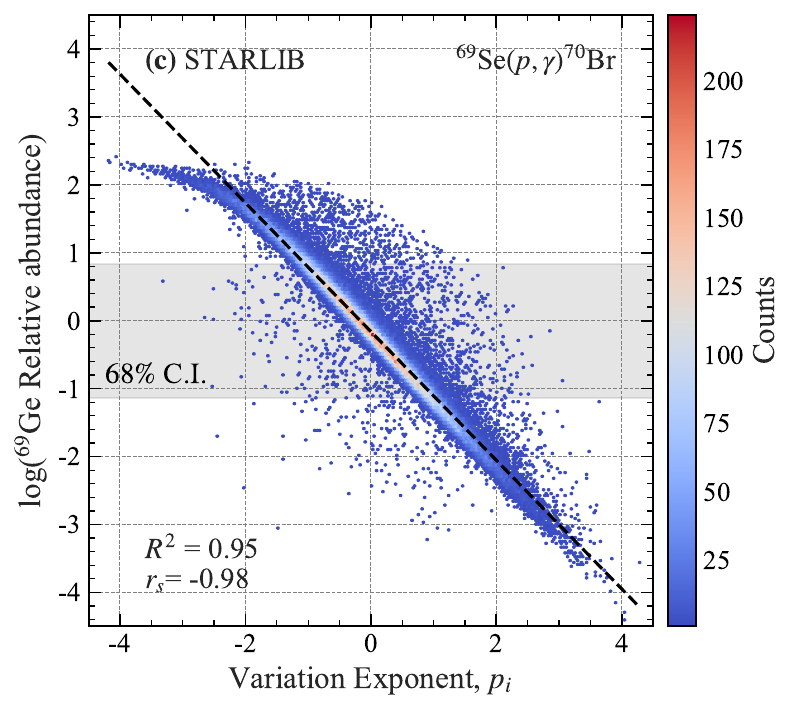}
\caption{\label{fig:Ge69}Effect of $^{69}$Se$(p, \gamma)^{70}$Br rate variations on the normalized abundance of $^{69}$Ge for Model K04 in three sampling schemes . The reaction shows a strong negative correlation with $^{69}$Ge. Compared to $f.u. = \sqrt{10}$ and STARLIB, the linear regression $R^2$ (black dashed line) and absolute Spearman correlation coefficient $|r_s|$ are smaller for $f.u. = 10$, while the corresponding abundance variation (gray band) is larger.}
\end{figure}

Figure~\ref{fig:Ge69} compares the correlations between variations in the $^{69}$Se$(p, \gamma)^{70}$Br rate and $^{69}$Ge abundance for the three sampling schemes. In particular, Figure~\ref{fig:Ge69}(b) shows a larger variation in the $^{69}$Ge abundance, with numerous discrete points distributed on both sides of the central region. Figure~\ref{fig:Ge69}(a) is consistent with Figure~8 of~\citet{Parikh2008}. The $^{69}\mathrm{Ge}$ abundance variations within the estimated uncertainty ranges are largest for $f.u. = 10$ (84th / 16th = 116.7), compared to $f.u. = \sqrt{10}$ (97.5th / 2.5th =95.6) and STARLIB (84th / 16th = 94.2). In contrast, the absolute Spearman coefficients $|r_s|$ are higher for $f.u. = \sqrt{10}$ (0.97) and STARLIB (0.98) than for $f.u. = 10$ (0.88).

\citet{Moreno2009} showed that the abundance distributions are close to normal for some isotopes with small uncertainties (e.g., $^{41}$Ca and $^{46}$Ti), while isotopic abundances with larger uncertainties (e.g., $^{15}$N and $^{60}$Ni) follow a lognormal distribution. In our analysis on a logarithmic scale, the distribution of isotopic abundances generally approximates either a normal or lognormal distribution. However, for $f.u. = 10$ and STARLIB, certain isotopes exhibit multi-peak structures, such as $^{55}$Co and $^{64}$Zn, whereas for $f.u. = \sqrt{10}$, no multi-peak structures are observed. We will discuss multi-peak structures of $^{55}$Co and $^{64}$Zn for $f.u.=10$ in Model K04, and the underlying mechanisms responsible for their formation, as follows.

\begin{figure}[!htbp]
   \centering
   \includegraphics[width=\columnwidth]{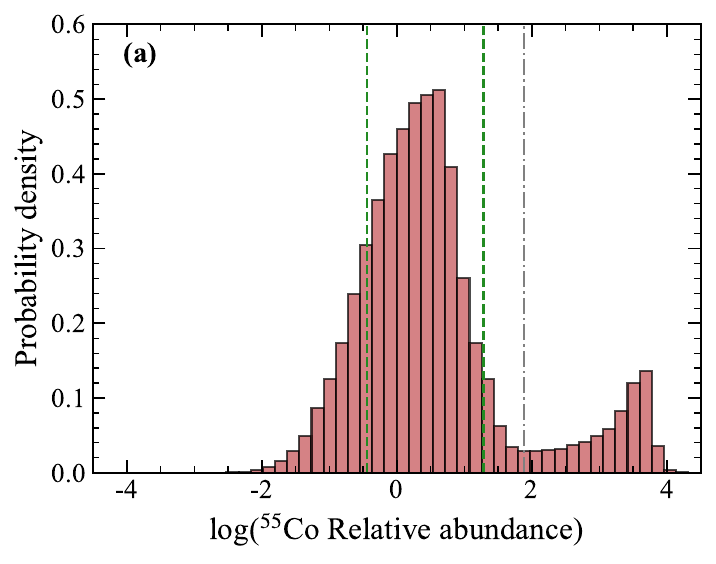}\\
   \includegraphics[width=\columnwidth]{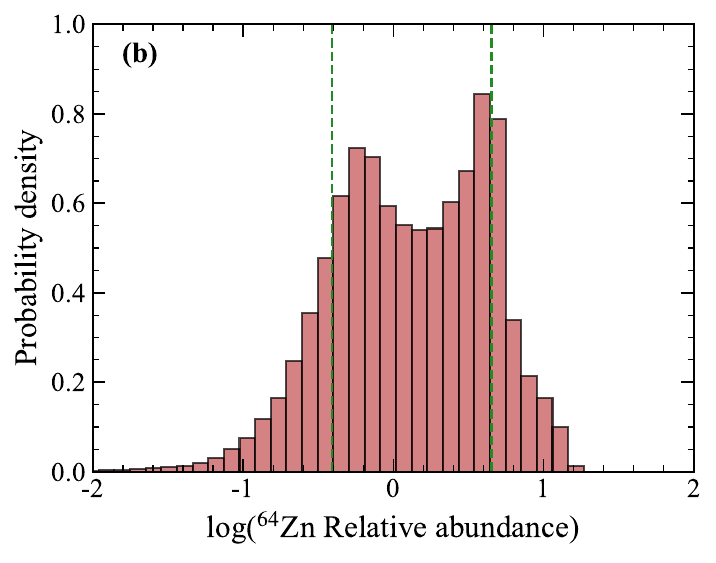} 
   \caption{\label{fig:Co55_and_Zn64}Histograms of the logarithmic relative abundances for $^{55}$Co (a) and $^{64}$Zn (b) in Model K04 for $f.u.=10$. The gray dot-dashed line in the top panel indicates the location of minimum probability density between the two peaks. The green dashed lines indicate the 68\% C.I. in each panel.}
\end{figure}

\begin{figure}[!hbp]
	\centering
	\includegraphics[width=\columnwidth]{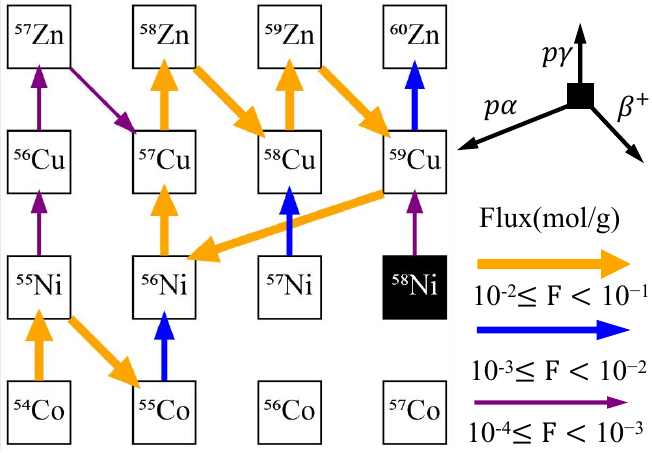}
	\caption{\label{fig:flow_co55}Integrated net fluxes for Model K04 at the end of the trajectory (the dashed-box region in Figure~\ref{fig:flow_standrates}). A representative Monte Carlo trial for $f.u.=10$ in which the ratio $^{59}$Cu$(p, \alpha)^{56}$Ni / $^{59}$Cu$(p, \gamma)^{60}$Zn is more than 100 times larger than the ratio of the median rates, illustrating a strong preference for the $(p,\alpha)$ channel, resulting in the high $^{55}$Co abundance peak. Flux strengths are indicated by arrow sizes and colors, with the corresponding legend shown on the right.}
\end{figure}

\begin{figure}[!htbp]
	\centering
	\includegraphics[width=\columnwidth]{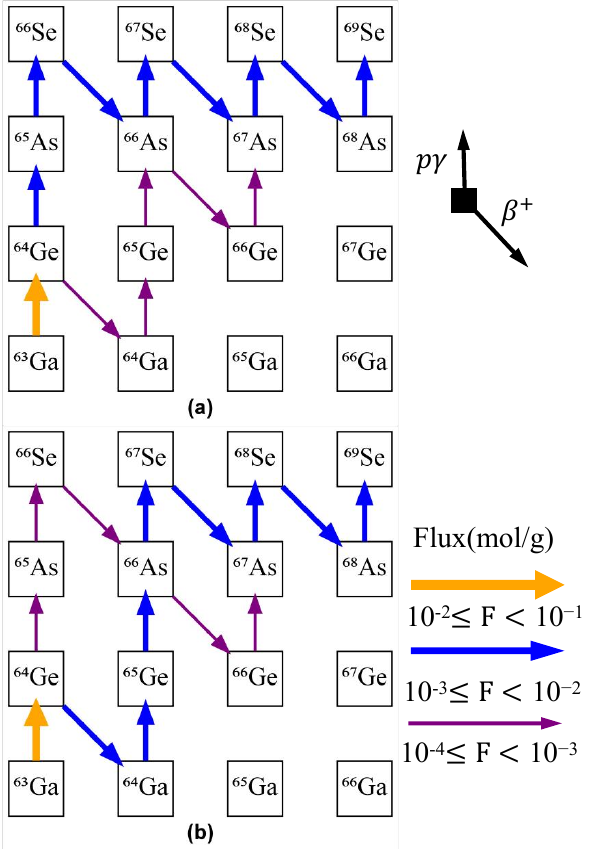}
	\caption{\label{fig:Zn64C}Integrated net fluxes for $f.u.=10$ in Model K04 at the end of the trajectory (the dot-dashed box region in Figure~\ref{fig:flow_standrates}). (a) A representative Monte Carlo trial corresponding to the peak at low $^{64}$Zn abundance. (b) A representative Monte Carlo trial corresponding to the peak at high $^{64}$Zn abundance. The two trials follow different dominant reaction paths, as indicated by the blue arrows. $^{64}$Ge is the waiting point in the network. Flux strengths are indicated by arrow sizes and colors, with the corresponding legend shown on the right.}
\end{figure}

Figure~\ref{fig:Co55_and_Zn64} shows the bimodal distributions of $^{55}$Co and $^{64}$Zn abundances. For $^{55}$Co (panel a), the high-abundance peak lies outside the 68\% C.I., and the integrated net fluxes of one representative Monte Carlo trial at this peak are shown in Figure~\ref{fig:flow_co55}. For $^{64}$Zn (panel b), both peaks are within the 68\% C.I., with the corresponding integrated net fluxes shown in Figure~\ref{fig:Zn64C}.

\begin{figure*}[!htbp]
	\centering
	\gridline{		
		\fig{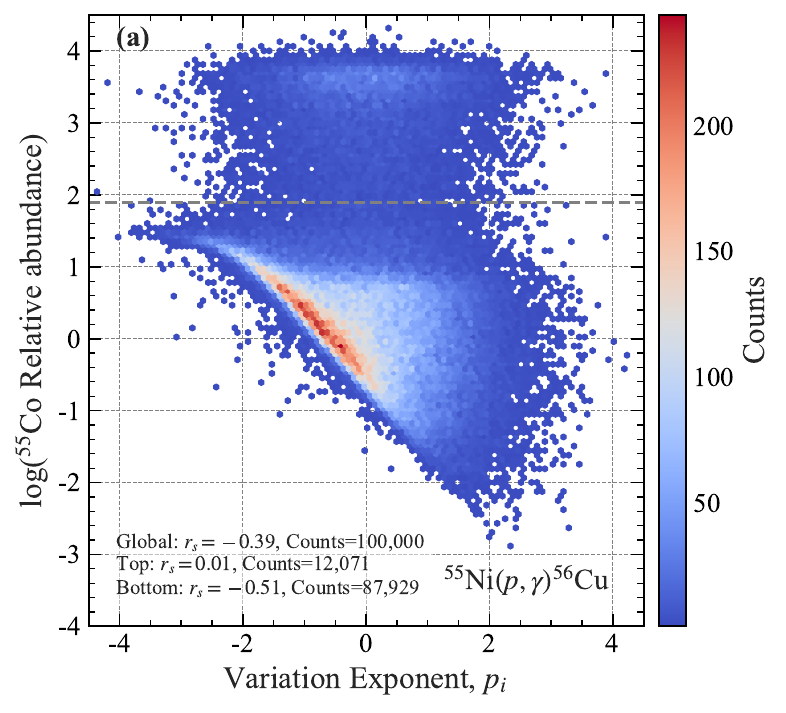}{0.5\textwidth}{}
		\fig{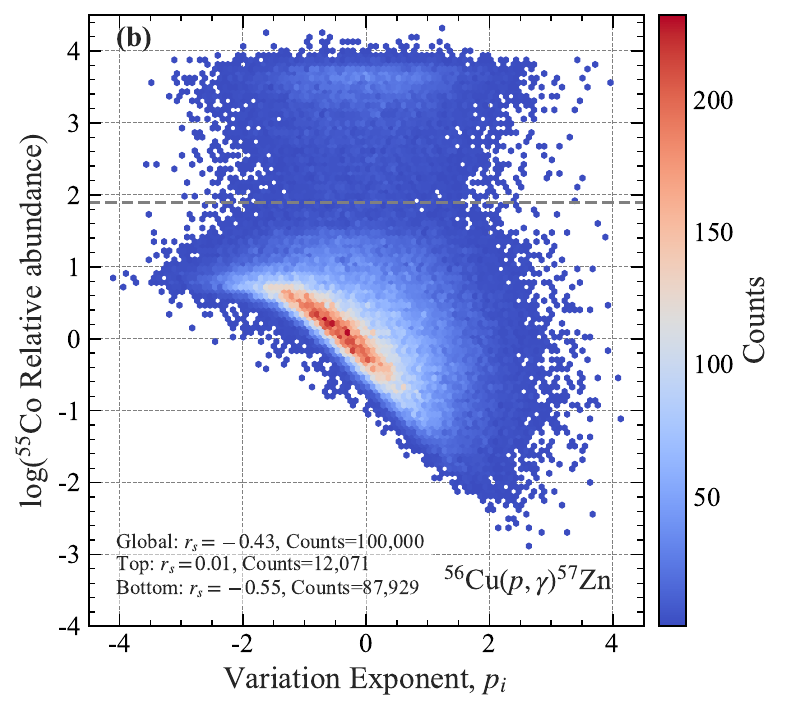}{0.5\textwidth}{}
	}
	\vspace{-30pt}
	\gridline{
		\fig{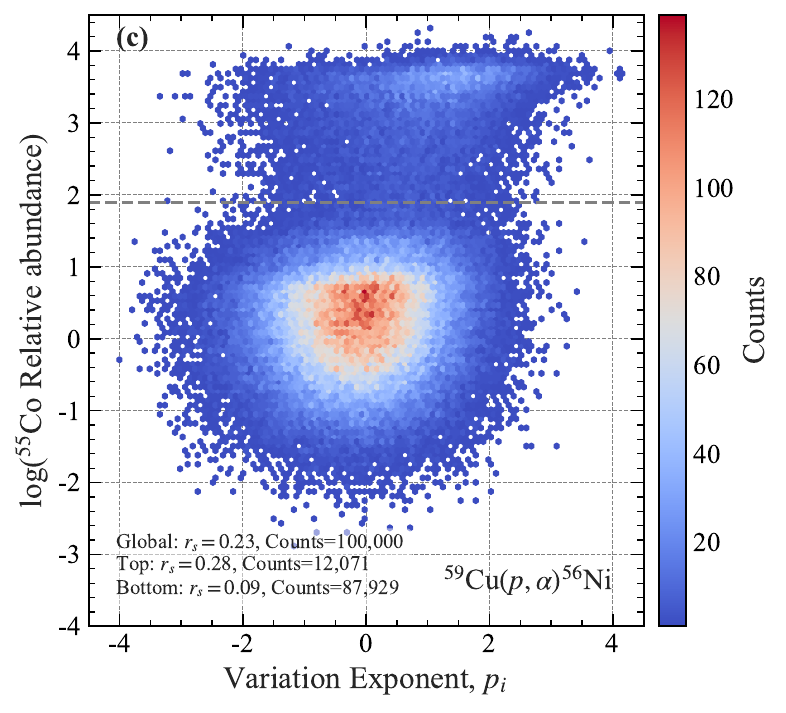}{0.5\textwidth}{}
		\fig{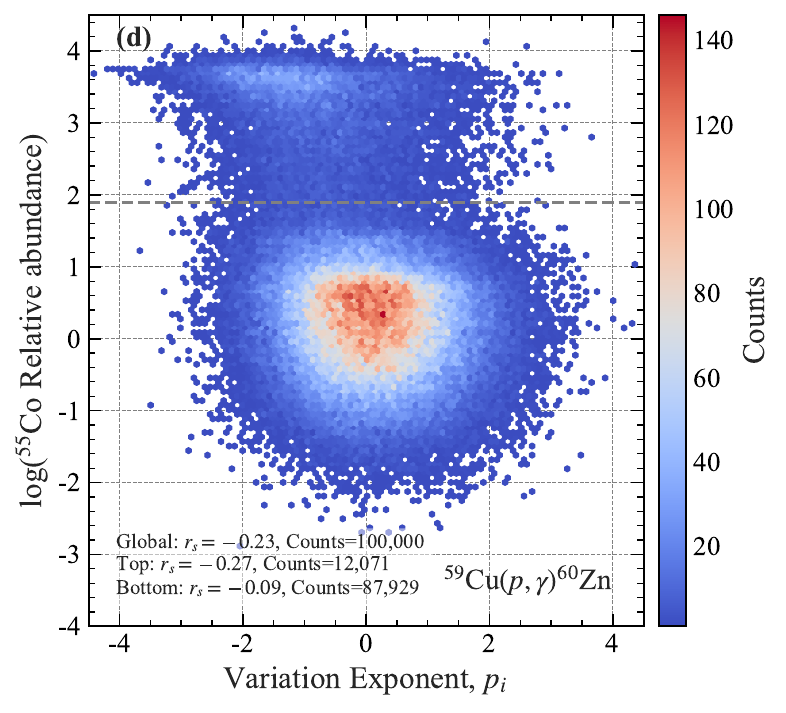}{0.5\textwidth}{}
	}
	\vspace{-20pt}
	\caption{\label{fig:Co55B}Similar to Figure~\ref{fig:Ge69}, but for the correlations between the abundance of $^{55}$Co and the reactions (a) $^{55}$Ni$(p, \gamma)^{56}$Cu, (b) $^{56}$Cu$(p, \gamma)^{57}$Zn, (c) $^{59}$Cu$(p, \alpha)^{56}$Ni, and (d) $^{59}$Cu$(p, \gamma)^{60}$Zn in Model K04 for $f.u = 10$. Gray dashed lines separate the primary (bottom) and secondary (top) clusters, corresponding to the minimum of the probability density of $^{55}$Co indicated by the gray dot-dashed line in Figure~\ref{fig:Co55_and_Zn64}(a). Each cluster is analyzed individually, with the Spearman correlation coefficient ($r_s$) and sample counts reported at the bottom of each panel.}
\end{figure*}

For the $^{55}$Co abundance, correlation analysis identifies four reactions as the main contributors to its abundance pattern: $^{55}$Ni$(p, \gamma)^{56}$Cu ($r_s=-0.39$),  $^{56}$Cu$(p, \gamma)^{57}$Zn ($r_s=-0.43$), $^{59}$Cu$(p, \alpha)^{56}$Ni($r_s=0.23$) and $^{59}$Cu$(p, \gamma)^{60}$Zn ($r_s=-0.23$). The effects of rate variations on the abundance of $^{55}$Co are shown in Figure~\ref{fig:Co55B}, where the abundances of $^{55}$Co form two distinct clusters, separated by the value corresponding to the minimum probability density between the two peaks in Figure~\ref{fig:Co55_and_Zn64}(a). Within the primary clusters (Figure~\ref{fig:Co55B}a–b), the abundance of $^{55}$Co shows stronger correlations with $^{55}$Ni$(p, \gamma)^{56}$Cu ($r_s=-0.51$) and $^{56}$Cu$(p, \gamma)^{57}$Zn ($r_s=-0.55$) compared to the global cluster ($r_s=-0.39$ and $r_s=-0.43$), while correlations in the secondary clusters are weak. In contrast, within the secondary clusters (Figure~\ref{fig:Co55B}c–d), $^{55}$Co exhibits slightly stronger correlations with $^{59}$Cu$(p, \alpha)^{56}$Ni ($r_s=0.28$) and $^{59}$Cu$(p, \gamma)^{60}$Zn ($r_s=-0.27$) than in the global cluster ($r_s=0.23$ and $r_s=-0.23$), while the correlations in the primary clusters remain weak.

\begin{figure}[!htbp]
\centering
\includegraphics[width=\columnwidth]{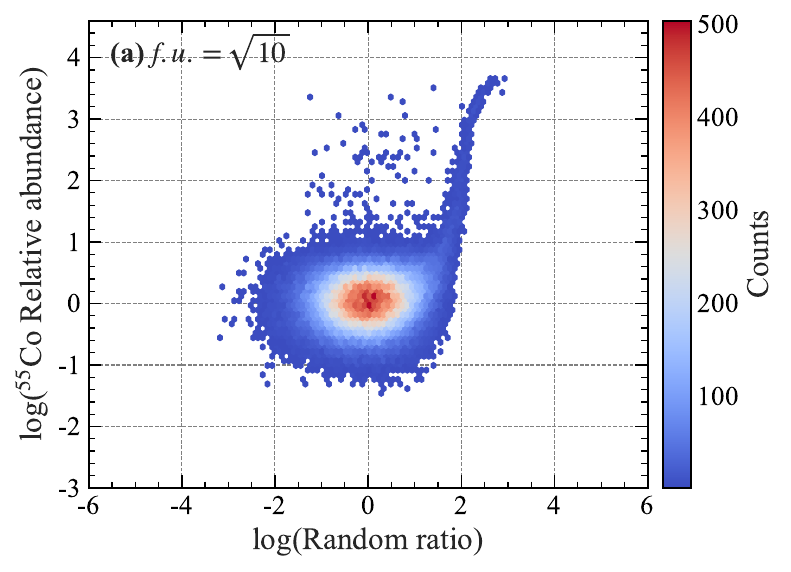}\\
\includegraphics[width=\columnwidth]{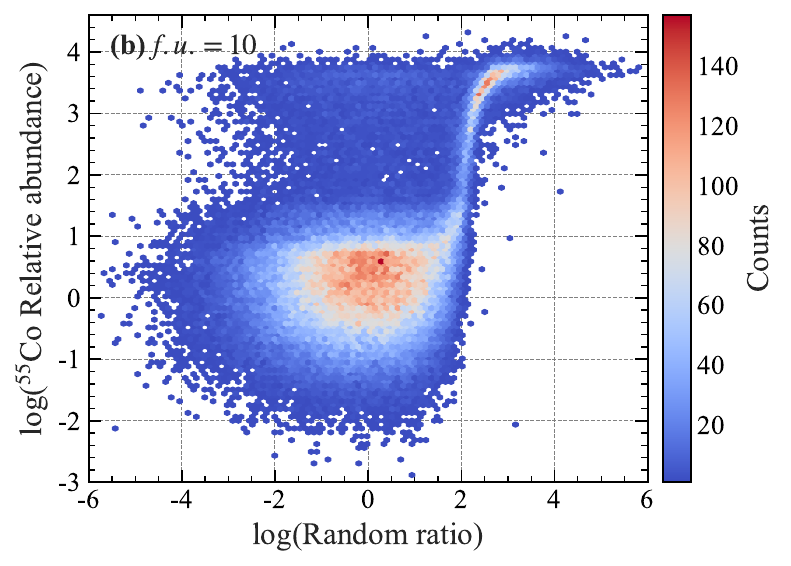}
\caption{\label{fig:Co55C}Effect of the ratio $^{59}$Cu$(p, \alpha)^{56}$Ni/$^{59}$Cu$(p, \gamma)^{60}$Zn relative to its median value, on the abundance of $^{55}$Co in Model K04. (a) $f.u.=\sqrt{10}$: the distribution of $^{55}$Co remains single-peaked with a slight tail. (b) $f.u.=10$: the ratio clearly separates trials leading to the high- and low-abundance peaks, illustrating the formation of the bimodal distribution.} 
\end{figure}

Figure~\ref{fig:Co55C}(b) further confirms that the bimodal distribution of $^{55}$Co mainly arises from the competition between $^{59}$Cu$(p, \gamma)^{60}$Zn and $^{59}$Cu$(p, \alpha)^{56}$Ni. The ratio of their assigned variation factors on median rates clearly separates the trials leading to high- and low-abundance peaks. For $f.u. = \sqrt{10}$ (Figure~\ref{fig:Co55C}(a)), the $^{55}$Co distribution remains single-peaked with a slight tail. The observed separation originates from the competition between the $p$-capture and $\alpha$-capture on $^{59}$Cu. When the ratio exceeds a threshold, the abundance distribution may depart from the
standard single-peak structure and exhibit a secondary peak, becoming increasingly likely with larger Monte Carlo perturbations. 

Figure~\ref{fig:flow_co55} presents the integrated net fluxes in the NiCu region for the high $^{55}$Co abundance peak, in which the ratio $^{59}$Cu$(p, \alpha)^{56}$Ni / $^{59}$Cu$(p, \gamma)^{60}$Zn relative to its median value exceeds 100, indicating a strong preference for the $(p,\alpha)$ channel. The increased flow of $(p,\alpha)$ strengthens the NiCu cycle, thereby accelerating hydrogen consumption and suppressing $p$-capture channel on $^{55}$Ni. more $^{55}$Ni material will decays into $^{55}$Co and $p$-capture channel on $^{55}$Co will be also suppressed due to less hydrogen. As a result, more material accumulates on $^{55}$Co, producing its bimodal abundance distribution and simultaneously reducing the flow to heavier Zn isotopes. Although the abundance of $^{55}$Co shows weak correlations with $^{59}$Cu$(p, \alpha)^{56}$Ni and $^{59}$Cu$(p, \gamma)^{60}$Zn under large perturbations, which is below the threshold of $r_s = 0.5$ used to define key reactions in this study, these reactions may nevertheless play a non-negligible role in shaping the final $^{55}$Co abundance. These results highlight the sensitivity of isotopic abundances to the $^{59}$Cu$(p,\alpha)/(p, \gamma)$ branching ratio and the role of the NiCu cycle in shaping the final composition.

\begin{figure}[!htbp]
	\includegraphics[width=\columnwidth]{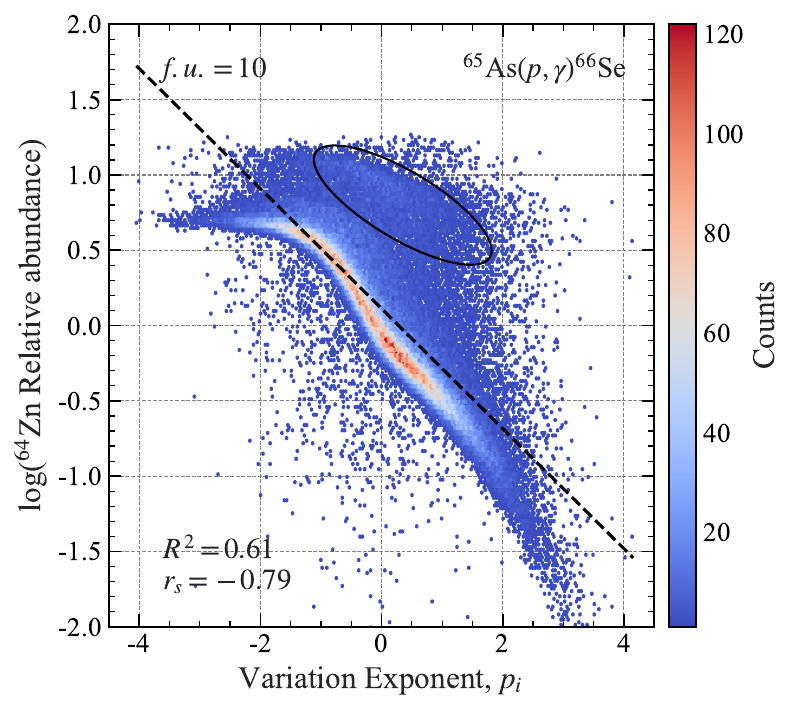}
	\caption{\label{fig:Zn64A}Similar to Figure~\ref{fig:Ge69}, but showing a good correlation between $^{65}$As$(p, \gamma)^{66}$Se and $^{64}$Zn for $f.u.=10$ in Model K04. $^{59}$Cu$(p, \alpha)^{56}$Ni ($r_s=0.15$) and $^{59}$Cu$(p, \gamma)^{60}$Zn ($r_s=-0.15$) show weak correlations with $^{64}$Zn, but leading to a sparse cluster in the region of the black ellipse.}
\end{figure}

For the $^{64}$Zn abundance, the bimodal distribution is mainly caused by the single reaction $^{65}$As$(p, \gamma)^{66}$Se ($r_s=-0.79$). $^{59}$Cu$(p, \alpha)^{56}$Ni ($r_s=0.15$) and $^{59}$Cu$(p, \gamma)^{60}$Zn ($r_s=-0.15$) also show weak correlations with $^{64}$Zn, leading to a sparse cluster in the black ellipse region as shown in Figure~\ref{fig:Zn64A}. An additional 100,000 Monte Carlo trials were performed, varying only the $^{65}$As$(p, \gamma)^{66}$Se rate, and the results are consistent. 

The reaction paths reveal the origin of the bimodal distribution. Figure~\ref{fig:Zn64C} shows the diagrams of the integrated net fluxes for two representative Monte Carlo trials at the end of the trajectory. The reaction flux inevitably passes through the waiting point $^{64}$Ge ($T_{1/2} = 63.7~\mathrm{s}$)~\citep{Lam2022c,Zhou2023}. In Figure~\ref{fig:Zn64C}(a), the first peak of $^{64}$Zn abundance is associated with higher $^{65}$As$(p, \gamma)^{66}$Se rates. The reaction flow mainly breaks out of the ZnGa cycle through the $^{63}$Ga$(p, \gamma)^{64}$Ge$(p, \gamma)^{65}$As path. This transfers more material through the region above Se, as reported by \citet{Lam2022b,Lam2025}, and results in lower abundances of $^{64}$Ge and $^{64}$Ga. In Figure~\ref{fig:Zn64C}(b), the second peak of $^{64}$Zn abundance is associated with lower $^{65}$As$(p, \gamma)^{66}$Se rates. The breakout occurs primarily through the alternative path $^{64}$Ga$(p, \gamma)^{65}$Ge$(p, \gamma)^{66}$As, leaving more material of $^{64}$Ge and $^{64}$Ga. As a result, the $\beta^+$-decays of $^{64}$Ge and $^{64}$Ga finally lead to a bimodal abundance distribution of $^{64}$Zn.

\section{\label{sec:Discussion and Conclusion}Discussion and Conclusion}
We systematically investigate the impact of nuclear reaction rate uncertainties on XRB nucleosynthesis using both temperature-independent and -dependent methods. Simulations were performed for three representative XRB trajectories (Models K04, S01, and S16) using the WinNet code, with additional cross-checks performed using the NucNet Tools~\citep{Meyer2007} and the SkyNet code~\citep{Lippuner2017}. Although there are slight differences in the final abundances calculated by these codes, the overall results are in good agreement. Our approach included 100,000 Monte Carlo trials per model run, enabling a comprehensive exploration of how final isotopic abundances are affected by variations across a large number of reactions.

Results from temperature-independent Monte Carlo simulations indicate that the choice of reaction rate factor uncertainty ($f.u.$) significantly affects both isotopic abundance uncertainties and the identification of key reactions. For modest perturbations ($f.u.=\sqrt{10}$), isotopic abundances exhibit limited variability, and the impact of individual reactions is more clearly reflected in the correlations, typically resulting in high Spearman coefficients. In contrast, with larger perturbations ($f.u.=10$), the uncertainties increase substantially, and the coupling among multiple competing reactions becomes more prominent. This reduces the linear correlation between individual reaction rates and final abundances, resulting in lower Spearman correlation coefficients.

Large perturbations also lead to greater variability and, in some cases, multi-peak abundance distributions. These multi-peak structures can arise not only from coupled reaction effects but also, in certain cases, from a single reaction, exemplified by $^{55}$Co and $^{64}$Zn. In contrast, smaller perturbations lead to less variability and predominantly single-peaked distributions, consistent with previous studies~\citep{Parikh2008,Moreno2009}. However, it should be noted that the pronounced effects associated with the ratio between the $^{59}$Cu$(p,\gamma)$ and $^{59}$Cu$(p,\alpha)$ reaction rates are obtained under the assumption that the uncertainties of these two rates are uncorrelated. Accounting for possible correlations between the two rates may make such extreme ratios either less or more likely, potentially modifying the resulting abundance distributions.

Temperature-dependent Monte Carlo simulations using STARLIB further incorporate more realistic, temperature-dependent rate uncertainties. We recommend using the temperature-dependent Monte Carlo method with STARLIB as the primary approach, while the temperature-independent Monte Carlo method with modest perturbations can be used for supplementary analysis. In all models, the key reactions identified align in large part with previous studies~\citep{Parikh2008, Moreno2009, Cyburt2016}. These findings offer insights for future experimental and theoretical efforts to reduce reaction rate uncertainties. In particular, reactions that show consistent correlations across multiple models and simulation methods should be prioritized for improved measurements.

When using the temperature-independent Monte Carlo method, $f.u.$ should be appropriately selected and assigned with a confidence range; each reaction should have its own uncertainty limit rather than applying the same uncertainty to all reactions. The temperature-dependent Monte Carlo method using STARLIB provides a more reliable, physically motivated assessment of abundance uncertainties than the temperature-independent method. Multi-peak abundance distributions naturally emerge from competing reaction pathways under large perturbations, emphasizing the non-linear, non-monotonic nature of XRB nucleosynthesis and warranting further study.

Overall, this work advances our understanding of XRB nucleosynthesis and highlights the importance of incorporating realistic, temperature-dependent reaction rate uncertainties. Future studies could extend this approach to 1D multizone hydrodynamic models, thereby providing an even more comprehensive assessment of reaction rate uncertainties in XRB nucleosynthesis. 

\begin{acknowledgments}
The authors thank the reviewer for her/his valuable comments on
the submitted manuscript. The authors thank H. Schatz for providing
the XRB trajectories used in this study. The authors acknowledge support from the National Natural Science Foundation of China (No. 12175152), LingChuang Research Project of the China National Nuclear Corporation (Nos. 2025-084, 2024-065, 2024-068).
\end{acknowledgments}

\software{\texttt{IPython}~\citep{ipython}, \texttt{Jupyter}~\citep{jupyter}, \texttt{matplotlib}~\citep{matplotlib}, \texttt{numpy}~\citep{numpy}, \texttt{scipy}~\citep{2020SciPy-NMeth}, \texttt{scikit-learn}~\citep{scikit-learn}, \texttt{seaborn}~\citep{Waskom2021}, \texttt{NucNet Tools}~\citep{Meyer2007}, \texttt{SkyNet}~\citep{Lippuner2017}, \texttt{WinNet}~\citep{Winteler2012,Reichert2023}}

\appendix

\startlongtable
\begin{deluxetable*}{ccccccccccccc}
\setlength{\tabcolsep}{3pt}
\tablecaption{\label{tab:Uncertainty Table}Isotopic abundance variations for the studied models in the temperature-dependent (STARLIB: 84\textsuperscript{th}/16\textsuperscript{th} percentile ratio) and -independent Monte Carlo simulations ($f.u.= \sqrt{10}$: 97.5\textsuperscript{th}/2.5\textsuperscript{th} percentile ratio, $f.u.= 10$: 84\textsuperscript{th}/16\textsuperscript{th} percentile ratio).
Isotopic abundance variations ($X_{\rm base} \ge 10^{-6}$) are at least a factor of 2. The full table of all isotopes satisfying these criteria is available online.
}
\tablehead{
	\colhead{} &
	\multicolumn{3}{c}{Model K04} &
	\multicolumn{3}{c}{Model S01} &
	\multicolumn{3}{c}{Model S16} \\
	\cline{2-4}
	\cline{5-7}
	\cline{8-10}
	\colhead{Isotope} &
	\colhead{STARLIB} & \colhead{$f.u.=\sqrt{10}$} & \colhead{$f.u.=10$} &
	\colhead{STARLIB} & \colhead{$f.u.=\sqrt{10}$} & \colhead{$f.u.=10$} &
	\colhead{STARLIB} & \colhead{$f.u.=\sqrt{10}$} & \colhead{$f.u.=10$}
}
\startdata
$\isotope[1][]{H}$ & ... & 2.2 & 4.8 & 2.6 & 4708.6 & $>10^4$ & ... & ... & ... \\
$\isotope[14][]{N}$ & 107.6 & ... & ... & 131.2 & 752.8 & $>10^4$ & 3.5 & ... & 2.0 \\
$\isotope[15][]{N}$ & 6.7 & 374.8 & 2411.2 & 6.6 & 1840.9 & $>10^4$ & 2.7 & 8.8 & 10.0 \\
$\isotope[18][]{F}$ & 2.5 & 41.0 & 235.7 & ... & 95.9 & $>10^4$ & 26.8 & 38.5 & 38.2 \\
$\isotope[20][]{Ne}$ & ... & ... & ... & ... & ... & ... & ... & 13.4 & 14.5 \\
$\isotope[21][]{Ne}$ & ... & 40.1 & 142.7 & ... & 36.7 & $>10^4$ & ... & 13.4 & 16.5 \\
$\isotope[22][]{Na}$ & 11.6 & 41.2 & 109.4 & 7.8 & ... & ... & 6.9 & 60.1 & 76.5 \\
$\isotope[23][]{Na}$ & ... & ... & ... & ... & 2.0 & $>10^4$ & 5.9 & 17.9 & 24.2 \\
$\isotope[55][]{Co}$ & 23.1 & 26.5 & 53.1 & 24.0 & ... & ... & 82.2 & 38.5 & 49.1 \\
$\isotope[56][]{Ni}$ & 12.3 & 11.7 & 69.1 & 27.2 & 129.1 & 1172.5 & 277.3 & 175.0 & 397.1 \\
$\isotope[57][]{Ni}$ & 3.6 & 3.1 & 5.7 & 9.0 & 9.4 & 19.1 & 29.0 & 17.2 & 22.2 \\
$\isotope[58][]{Ni}$ & ... & ... & ... & 2.9 & 2.2 & 4.4 & 24.0 & 28.6 & 30.6 \\
$\isotope[59][]{Ni}$ & ... & ... & 2.9 & ... & 4.2 & 4.4 & 32.4 & 29.7 & 32.4 \\
$\isotope[60][]{Ni}$ & 47.0 & 94.7 & 250.4 & 34.9 & 248.3 & 313.6 & 24.8 & 44.2 & 44.9 \\
$\isotope[61][]{Ni}$ & ... & ... & ... & ... & ... & ... & 78.8 & ... & ... \\
$\isotope[62][]{Ni}$ & ... & ... & ... & ... & ... & ... & 37.4 & 60.2 & 58.4 \\
$\isotope[61][]{Cu}$ & 51.8 & 61.5 & 94.1 & 40.6 & 117.9 & 410.8 & 45.1 & 70.6 & 81.6 \\
$\isotope[63][]{Cu}$ & 2.9 & 3.3 & 4.6 & 7.7 & 11.6 & 11.6 & 36.0 & 21.8 & 24.3 \\
$\isotope[62][]{Zn}$ & 2.0 & ... & ... & 3.2 & 2.3 & 4.6 & 39.8 & 85.4 & 80.9 \\
$\isotope[64][]{Zn}$ & 11.7 & 10.0 & 11.5 & 2.4 & 2.2 & 4.5 & 3.1 & ... & ... \\
$\isotope[65][]{Zn}$ & 79.7 & 130.4 & 120.5 & 102.6 & 139.2 & 229.0 & 12.7 & 11.6 & 12.6 \\
$\isotope[66][]{Ga}$ & ... & ... & ... & ... & ... & ... & 15.7 & 16.0 & 16.6 \\
$\isotope[67][]{Ga}$ & 15.4 & 30.8 & 27.1 & 29.6 & 57.3 & 61.6 & 35.1 & 28.9 & 34.3 \\
$\isotope[66][]{Ge}$ & 23.1 & 37.6 & 32.3 & 19.6 & 37.0 & 47.3 & 20.1 & 20.1 & 20.7 \\
$\isotope[68][]{Ge}$ & ... & ... & ... & ... & ... & 3.7 & 4.8 & 2.3 & 2.7 \\
$\isotope[69][]{Ge}$ & 94.2 & 95.6 & 116.7 & 88.0 & 116.0 & 106.5 & 22.8 & 14.7 & 17.5 \\
\enddata
\end{deluxetable*}

\startlongtable
\begin{deluxetable*}{cccccccccccccc}
\setlength{\tabcolsep}{3pt}
\tablecaption{\label{tab:Key Reaction Table}The Spearman correlation coefficients $r_s$ between reactions and isotopes for the studied models in the temperature-dependent (STARLIB) and -independent Monte Carlo simulations ($f.u.=\sqrt{10}$ and $f.u.= 10$). Reactions with $|r_s| \ge 0.50$ that affect isotopic abundance variations ($X_{\rm base} \ge 10^{-6}$) by at least a factor of 2 are reported. The full table of all reactions satisfying these criteria is available online. 
}
\tablehead{
	\colhead{} &
	\colhead{} &
	\multicolumn{3}{c}{Model K04} &
	\multicolumn{3}{c}{Model S01} &
	\multicolumn{3}{c}{Model S16} \\
	\cline{3-5}
	\cline{6-8}
	\cline{9-12}
	\colhead{Reaction} &
	\colhead{Isotope} &
	\colhead{STARLIB} & \colhead{$f.u.=\sqrt{10}$} & \colhead{$f.u.=10$} &
	\colhead{STARLIB} & \colhead{$f.u.=\sqrt{10}$} & \colhead{$f.u.=10$} &
	\colhead{STARLIB} & \colhead{$f.u.=\sqrt{10}$} & \colhead{$f.u.=10$}
}
\startdata
$\isotope[14][]{O}(\alpha,p)\isotope[17][]{F}$&$\isotope[14][]{N}$ & $-0.97$ & ... & ... & $-0.92$ & $-0.95$ & $-0.57$ & $-0.98$ & ... & $-0.85$ \\
$\isotope[15][]{O}(\alpha,\gamma)\isotope[19][]{Ne}$&$\isotope[15][]{N}$ & $-0.85$ & $-0.75$ & $-0.60$ & $-0.88$ & $-0.80$ & ... & $-0.56$ & $-0.94$ & $-0.82$ \\
$\isotope[18][]{Ne}(\alpha,p)\isotope[21][]{Na}$&$\isotope[18][]{F}$ & $-0.96$ & $-0.99$ & $-0.76$ & ... & $-0.95$ & $-0.56$ & ... & ... & ... \\
&$\isotope[21][]{Ne}$ & ... & $-0.99$ & $-0.77$ & ... & $-0.95$ & $-0.57$ & ... & ... & ... \\
$\isotope[55][]{Ni}(p,\gamma)\isotope[56][]{Cu}$&$\isotope[55][]{Co}$ & $-0.80$ & ... & ... & $-0.81$ & ... & ... & ... & ... & ... \\
$\isotope[56][]{Ni}(p,\gamma)\isotope[57][]{Cu}$&$\isotope[56][]{Ni}$ & $-0.90$ & $-0.92$ & $-0.67$ & $-0.84$ & $-0.84$ & $-0.50$ & $-0.94$ & $-0.91$ & $-0.85$ \\
$\isotope[57][]{Ni}(p,\gamma)\isotope[58][]{Cu}$&$\isotope[57][]{Ni}$ & ... & ... & ... & ... & ... & ... & $-0.82$ & $-0.69$ & $-0.70$ \\
&$\isotope[58][]{Ni}$ & ... & ... & ... & ... & ... & ... & ... & $0.55$ & ... \\
$\isotope[58][]{Ni}(p,\gamma)\isotope[59][]{Cu}$&$\isotope[58][]{Ni}$ & ... & ... & ... & ... & ... & ... & $-0.55$ & ... & ... \\
$\isotope[56][]{Cu}(p,\gamma)\isotope[57][]{Zn}$&$\isotope[55][]{Co}$ & ... & $-0.72$ & ... & ... & ... & ... & ... & ... & ... \\
$\isotope[57][]{Cu}(p,\gamma)\isotope[58][]{Zn}$&$\isotope[57][]{Ni}$ & $-0.78$ & $-0.76$ & $-0.55$ & $-0.80$ & $-0.78$ & ... & ... & ... & ... \\
$\isotope[58][]{Cu}(p,\gamma)\isotope[59][]{Zn}$&$\isotope[58][]{Ni}$ & ... & ... & ... & $-0.72$ & $-0.56$ & ... & ... & ... & $-0.57$ \\
$\isotope[59][]{Cu}(p,\gamma)\isotope[60][]{Zn}$&$\isotope[59][]{Ni}$ & ... & ... & $-0.66$ & ... & $-0.79$ & $-0.53$ & $-0.82$ & $-0.84$ & $-0.87$ \\
$\isotope[60][]{Zn}(p,\gamma)\isotope[61][]{Ga}$&$\isotope[60][]{Ni}$ & ... & ... & ... & ... & ... & ... & $-0.59$ & $-0.66$ & $-0.62$ \\
$\isotope[61][]{Zn}(p,\gamma)\isotope[62][]{Ga}$&$\isotope[61][]{Cu}$ & ... & ... & ... & ... & ... & ... & ... & $-0.55$ & $-0.53$ \\
$\isotope[62][]{Zn}(p,\gamma)\isotope[63][]{Ga}$&$\isotope[62][]{Zn}$ & ... & ... & ... & ... & ... & ... & ... & $-0.55$ & $-0.55$ \\
$\isotope[65][]{Zn}(p,\gamma)\isotope[66][]{Ga}$&$\isotope[66][]{Ga}$ & ... & ... & ... & ... & ... & ... & ... & $0.62$ & $0.51$ \\
$\isotope[61][]{Ga}(p,\gamma)\isotope[62][]{Ge}$&$\isotope[60][]{Ni}$ & $-0.92$ & $-0.93$ & $-0.69$ & $-0.78$ & $-0.81$ & $-0.50$ & $-0.70$ & $-0.65$ & $-0.57$ \\
&$\isotope[61][]{Ni}$ & ... & ... & ... & ... & ... & ... & $-0.57$ & ... & ... \\
&$\isotope[62][]{Ni}$ & ... & ... & ... & ... & ... & ... & $-0.56$ & $-0.61$ & $-0.56$ \\
&$\isotope[61][]{Cu}$ & $-0.93$ & $-0.97$ & $-0.70$ & $-0.81$ & $-0.89$ & $-0.55$ & $-0.66$ & $-0.63$ & $-0.62$ \\
&$\isotope[62][]{Zn}$ & ... & ... & ... & ... & $0.54$ & ... & $-0.54$ & $-0.56$ & $-0.52$ \\
&$\isotope[64][]{Zn}$ & ... & ... & ... & ... & ... & ... & $0.77$ & ... & ... \\
&$\isotope[68][]{Ge}$ & ... & ... & ... & ... & ... & ... & $0.76$ & ... & ... \\
$\isotope[62][]{Ga}(p,\gamma)\isotope[63][]{Ge}$&$\isotope[62][]{Zn}$ & $-0.70$ & ... & ... & $-0.68$ & ... & ... & ... & ... & ... \\
$\isotope[63][]{Ga}(p,\gamma)\isotope[64][]{Ge}$&$\isotope[63][]{Cu}$ & $-0.57$ & $-0.94$ & $-0.78$ & $-0.72$ & $-0.89$ & $-0.63$ & $-0.60$ & $-0.55$ & $-0.64$ \\
$\isotope[64][]{Ga}(p,\gamma)\isotope[65][]{Ge}$&$\isotope[65][]{Zn}$ & ... & ... & ... & ... & ... & ... & $0.80$ & $0.92$ & $0.77$ \\
&$\isotope[66][]{Ga}$ & ... & ... & ... & ... & ... & ... & ... & $0.55$ & ... \\
$\isotope[65][]{Ge}(p,\gamma)\isotope[66][]{As}$&$\isotope[65][]{Zn}$ & $-0.83$ & $-0.86$ & $-0.78$ & $-0.89$ & $-0.89$ & $-0.65$ & ... & ... & ... \\
&$\isotope[66][]{Ge}$ & ... & ... & ... & ... & ... & ... & ... & $0.53$ & ... \\
$\isotope[66][]{Ge}(p,\gamma)\isotope[67][]{As}$&$\isotope[66][]{Ge}$ & ... & ... & ... & $-0.64$ & $-0.65$ & ... & ... & $-0.55$ & ... \\
$\isotope[65][]{As}(p,\gamma)\isotope[66][]{Se}$&$\isotope[64][]{Zn}$ & $-0.94$ & $-0.97$ & $-0.79$ & $-0.74$ & $-0.70$ & $-0.54$ & ... & ... & ... \\
&$\isotope[67][]{Ga}$ & $-0.57$ & $-0.63$ & ... & ... & ... & ... & ... & ... & ... \\
&$\isotope[66][]{Ge}$ & ... & $-0.51$ & ... & ... & ... & ... & ... & ... & ... \\
&$\isotope[68][]{Ge}$ & ... & ... & ... & ... & ... & ... & ... & $0.75$ & $0.56$ \\
$\isotope[66][]{As}(p,\gamma)\isotope[67][]{Se}$&$\isotope[66][]{Ge}$ & $-0.70$ & $-0.71$ & $-0.55$ & ... & ... & ... & ... & ... & ... \\
$\isotope[67][]{As}(p,\gamma)\isotope[68][]{Se}$&$\isotope[67][]{Ga}$ & $-0.59$ & $-0.63$ & $-0.66$ & $-0.86$ & $-0.89$ & $-0.68$ & $-0.74$ & $-0.80$ & $-0.80$ \\
$\isotope[68][]{As}(p,\gamma)\isotope[69][]{Se}$&$\isotope[69][]{Ge}$ & ... & ... & ... & ... & ... & ... & $0.66$ & $0.81$ & $0.65$ \\
$\isotope[69][]{Se}(p,\gamma)\isotope[70][]{Br}$&$\isotope[69][]{Ge}$ & $-0.98$ & $-0.97$ & $-0.88$ & $-0.92$ & $-0.90$ & $-0.71$ & ... & ... & ... \\
\enddata
\end{deluxetable*}

\bibliography{reference}{}
\bibliographystyle{aasjournalv7}
\end{document}